\documentclass[a4paper,11pt]{article}
\usepackage{pos}

\usepackage{lipsum} 
\usepackage{subcaption}
\usepackage[capitalise]{cleveref}
\usepackage{siunitx}
\patchcmd{\thebibliography}
  {\labelsep}
  {\itemsep 0pt \labelsep}
  {}{}

\usepackage{comment}

\newcommand{\be}{\begin{equation}}
\newcommand{\ee}{\end{equation}}

\def\X2{Var($X_{\max})$}

\title{Monte Carlo parameter study for Seyfert AGN-starburst composite galaxies NGC1068 and NGC7469}
 \ShortTitle{Monte Carlo parameter study for NGC1068 and NGC7469}

\author*[a,b]{Silvia Salvatore}
\author[a,b]{Bj\"orn Eichmann}
\author[c]{Giacomo Sommani}
\author[d]{Santiago del Palacio}
\author[c]{Patrik M. Veres}
\author[a,b,d]{Julia Becker Tjus}

\affiliation[a]{Ruhr-Universit\"at Bochum, Theoretische Physik IV, Fakult\"at f\"ur Physik und Astronomie, Bochum, Germany}
\affiliation[b]{Ruhr Astroparticle and Plasma Physics Center (RAPP Center), Bochum, Germany}
\affiliation[c]{Ruhr-Universit\"at Bochum, Astronomical Institute, Fakult\"at f\"ur Physik und Astronomie, Bochum, Germany}
\affiliation[d]{Department of Space, Earth and Environment, Chalmers University of Technology, 412 96
Gothenburg, Sweden}

\emailAdd{silvia.salvatore@rub.de}

\abstract{

Seyfert-starburst composite galaxies host two promising phenomena of non-thermal high-energy radiation. In this regard the IceCube observation of high-energy neutrinos from the direction of the Seyfert-starburst composite galaxy NGC\,1068 is not surprising. More recently, another Seyfert-starburst composite galaxy, NGC\,7469, has shown hints for neutrino emission at even higher energies. Theoretical investigations could clarify that their so-called AGN corona is the most-likely origin of these neutrinos due to the need of being partially $\gamma$-ray opaque.

In this work, we present an updated version of our Seyfert-starburst composite model from 2022, that accounts for a proper treatment of the stochastic acceleration processes in the AGN corona and the secondary electrons and positrons from leptonic radiation processes. Moreover, we use a Markov Chain Monte Carlo (MCMC) approach to study the parameter space of these two potential high-energy neutrino sources under consideration of the given prior knowledge.

In the case of NGC\,1068, we can successfully explain its non-thermal observational features, where both its AGN corona and starburst ring are needed to account for the observations at high-energies. 
In the case of NGC\,7469, the high-energy signatures can only be explained assuming a small coronal radius and the including external $\gamma\gamma$-pair attenuation. 
In general, both sources exhibit a strong influence of the $\gamma$-ray opaqueness on the results, highlighting the need for an accurate treatment of the intrinsic coronal X-ray field and the spatial extent of the $\gamma$-ray production site.}

\ConferenceLogo{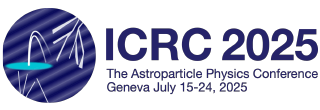}

\FullConference{%
39th International Cosmic Ray Conference (ICRC2025)\\
  15 - 24 July, 2025\\
 Geneva, Switzerland}

\begin{document}

\maketitle

\section{Introduction} \label{sec:intro}

Multimessenger astrophysics has significantly advanced since the beginning of the century, driven by the detection of high-energy charged particles and photons across various wavelengths. More recently, the field has been further enriched by astrophysical neutrino observations, such as those conducted by the IceCube telescope embedded in the Antarctic ice.
One very interesting multimessenger source type is Seyfert galaxies, which show electromagnetic emission from radio to X-ray or in some cases even $\gamma$-ray frequencies. However, this emission is in most frequency bands dominated by thermal radiation, and typically non-thermal emission becomes only dominant in the radio (even though they are still classified as radio weak) and/or $\gamma$-ray bands. 
They are active galactic nuclei (AGN) with a supermassive black hole and the surrounding dusty torus can absorb part of the emitted radiation by the core. Depending on the orientation of the torus relative to our line of sight, the Seyfert galaxies are classified by a number between 1 and 2. For Seyfert 2 galaxies the central emission is strongly attenuated by the torus, while Seyfert 1 are hardly affected by this. 

Most recently, the Seyfert-starburst composite galaxies NGC\,1068 (classified as Seyfert 2) and NGC\,7469 (classified as Seyfert 1 or 1.5) have been in the focus of astrophysical interest due to indications and hints, respectively, for high energy neutrino emission:
In case of NGC 1068, the IceCube Collaboration found \citep{icecube2022evidence}, with a global significance of $4.2\sigma$, a high-energy neutrino signal at TeV energies from the direction of NGC 1068 based on a dedicated catalog of 110 astronomical objects that are bright in the \textit{Fermi}-LAT energy band. 
In case of NGC\,7469, Sommani et al.~\citep{sommani2025two} found two IceCube real-time track alerts, out of just $\sim30$ per year from the direction of this source, and estimated a chance coincidence probability of 0.08\%~($3.2\,\sigma$).
The spectral dependence of the neutrino flux is unclear due to the limited number of neutrinos and their energy uncertainty. In contrast to NGC\,7469, NGC\,1068 also shows GeV $\gamma$-ray emission \citep{abdollahi2022incremental}.

A theoretical explanation of the whole non-thermal multimessenger emission of NGC\,1068 has been provided by a steady-state AGN-starburst composite model \cite{eichmann2022solving}, hereafter referred to as E+22 model. It has been shown that the observational features can be nicely explained if both emission sites are taken into account. A rather large CR pressure of $\gtrsim10\%$ of the thermal gas pressure is needed in the AGN corona, such as recently found by \cite{mutie2025consistent}. Moreover, the resulting spectral behavior of the non-thermal protons has been much softer than expected from stochastic diffusive acceleration. 

In this work, we present an updated version of the E+22 model with an improved treatment of (i) the stochastic acceleration process, as well as (ii) the source of secondary electrons and positrons in the AGN corona. Moreover, we analyse different $\gamma\gamma$-pair attenuation scenarios and use a Markov Chain Monte Carlo (MCMC) approach to study the parameter space of NGC\,1068 and NGC\,7469 based on their non-thermal emission features. 

The paper is structured as follows: In Sect.~\ref{sec:AGN-starburst-model}, we introduce the updated Seyfert-starburst composite model, and in Sect.~\ref{sec:methods} we explain the methods used for the comparison with the observational data of NGC\,1068 and NGC\,7469. In Sect.~\ref{sec:results} we show the results, before we conclude and discuss them in Sect.~\ref{sec:conclusions}.

\section{AGN-starburst composite model} \label{sec:AGN-starburst-model}
In the following, the AGN corona and the kiloparsec scale starburst ring of the Seyfert galaxies NGC\,1068 and NGC\,7469 are modelled independent of each other. Hereby, we assume a constant strength of the magnetic field (that is uniform on small scales and randomly orientated on significantly larger ones), as well as a thermal gas with a constant density $n$ and temperature $T_{\mathsf{gas}}$ in the different environments.

With respect to the non-thermal emission signatures the most relevant thermal photon targets are (i) the hard X-ray field for the AGN corona, and (ii) the infrared (IR) emission from the re-scattered starlight by dust grains in the starburst ring with an inner $R_{\rm in}^{\rm s}$ and outer $R_{\rm out}^{\rm s}$ radii (see E+22 for more details). In the case of NGC\,1068, observations indicate that $R_{\rm in}^{\rm s}=0.7\,\text{kpc}$ and $R_{\rm out}^{\rm s}=1.5\,\text{kpc}$. In both cases, for simplicity, we assume these thermal fields to be isotropic and stationary. Note that these assumptions have a significant impact on the $\gamma$-ray opacity of the corona, as discussed in more detail in Sect.~\ref{sec:conclusions}.
Moreover, both non-thermal emission sites are treated in a steady state and are considered to be spatially homogeneous for mathematical convenience.\\
In contrast to the E+22 model, the major changes are as follows:
\begin{enumerate}
    \item[(i)] Particle acceleration by stochastic diffusive acceleration (SDA) and energy losses occur simultaneously within the AGN corona.
    \item[(ii)] The secondary electrons $e^-$ and positrons $e^+$ originate not only from hadronic processes, which set the first generation of $e^\pm$ secondaries, but also from the leptonic radiation processes --- mainly synchrotron and IC --- starting from these secondaries.
\end{enumerate}

\begin{figure}[htbp]
    \centering
    \includegraphics[width=0.48\linewidth]{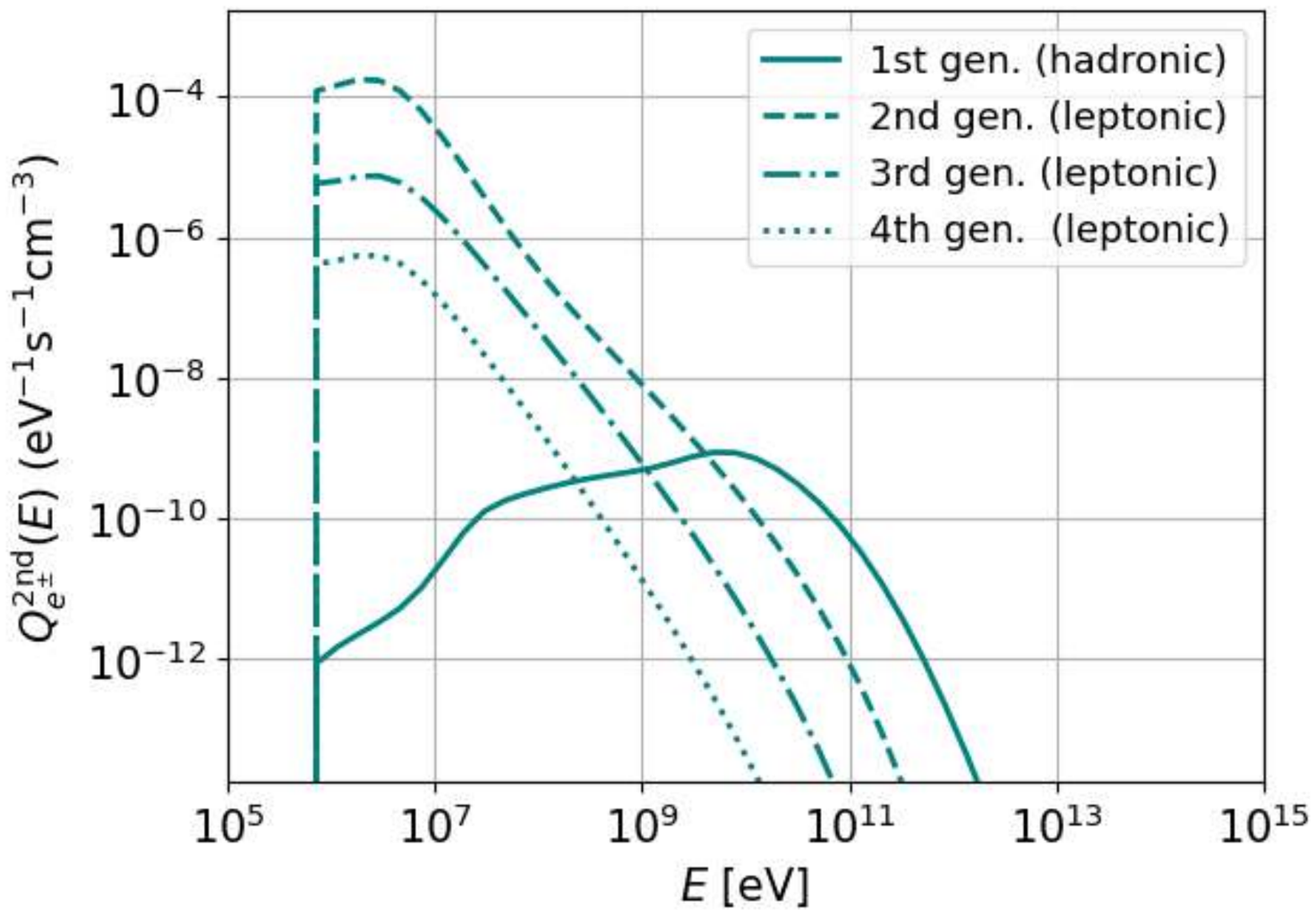}
    \includegraphics[width=0.48\linewidth]{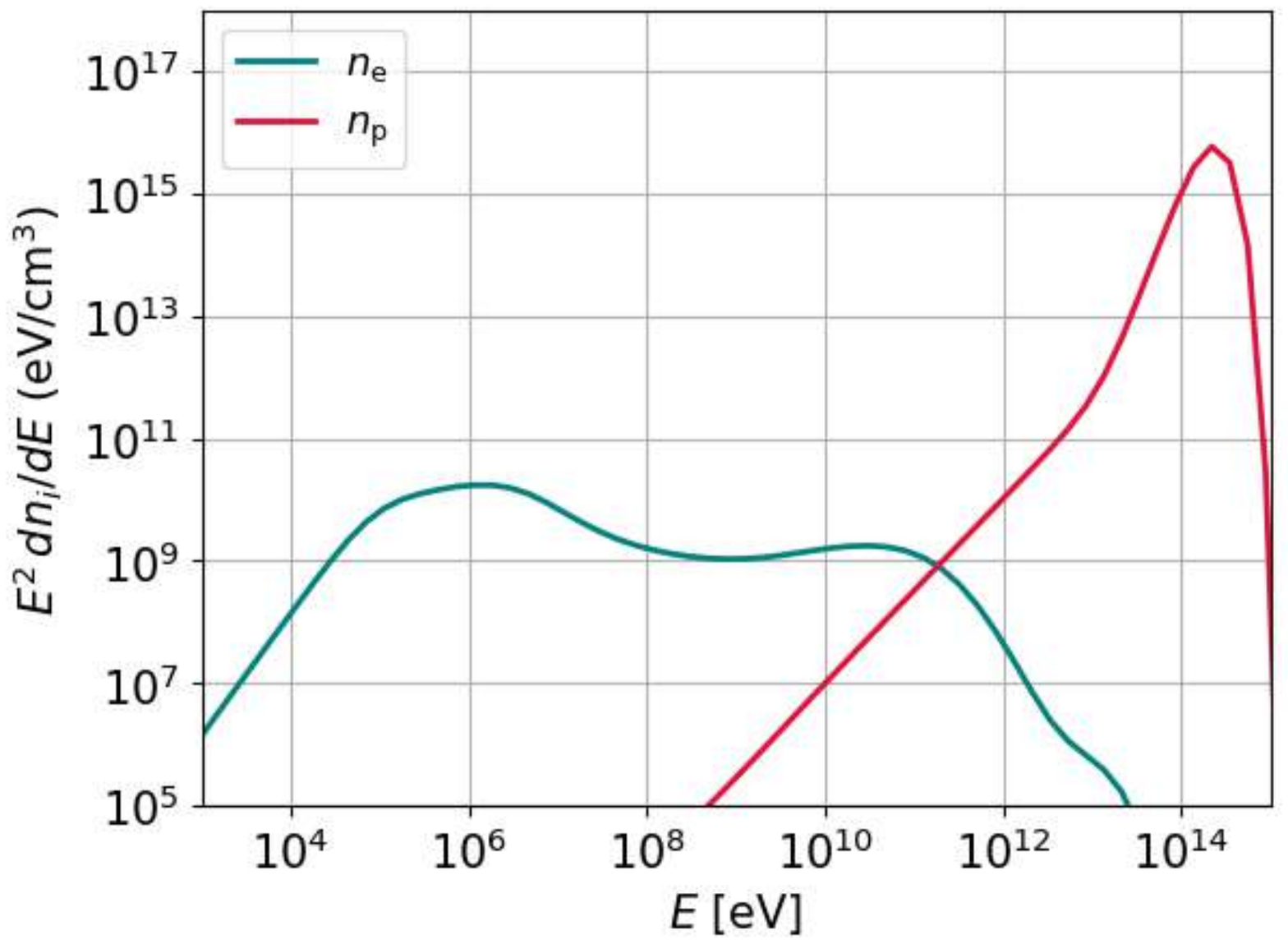}
    \caption{Relativistic particles energy distribution in the AGN corona. On the left, the secondary electrons and positrons source rate. On the right, the differential energy density of the electrons (including the secondary contribution) and protons.} 
    \label{fig:had+lept} 
\end{figure}

For the description of the transport equation, we follow Ref.~\cite{walter2025stochastic} and assume an assigned magnetic turbulence with a power spectrum $W(k)\propto k^{-\varkappa}$ and a strength $\eta^{-1}\equiv 8\pi \int W(k)\,\text{d}k/B^2$, where we used $\varkappa=5/3$ and $\eta^{-1}\simeq 0.8$ throughout this work. When the turbulent magnetic field is small compared to the regular magnetic field ($\eta^{-1} < 1$), the transport equation for the cosmic ray density $n_{i}$ in this region becomes
\begin{equation}
   -\frac{\partial}{\partial p}\left(D(p) \frac{\partial n_{i}}{\partial p}\right)+\frac{\partial}{\partial p}\left(\left(\frac{2 D(p)}{p}+\langle\dot{p}\rangle\right) n_{i}\right)+\frac{n_{i}}{\tau_{\mathrm{esc}}^{\rm c}(p)}=
    \begin{cases}
    Q_{\mathrm{p}}(p) \quad &\text{for} \quad {i=\mathrm{p}}\\
    Q_{\mathrm{e}}(p) + Q_{\mathrm{e}^{\pm}}^{2 \mathrm{nd}}(p)\quad &\text{for} \quad {i=\mathrm{e}}
    \end{cases}
    \label{eq:transportEQ}
\end{equation}
where
$D(p) \propto p^\varkappa$ 
is the momentum diffusion coefficient, describing the interaction rate with the magnetic turbulence. 
The continuous losses, due to interaction with the ambient gas and photon targets, are incorporated in the $\langle\dot{p}\rangle<0$ parameter; and the catastrophic losses, which are important for the resulting steady-state solution even in the case of ineffective particle escape on a timescale $\tau_{\rm esc}^{\rm c}(p)\gg |p/\langle\dot{p}\rangle|$, are also taken into account --- in contrast to a previous treatment \citep{eichmann2023disentangling}. All details on the energy losses, the escape and the acceleration timescales can be found in E+22. In addition to the primary source rate $Q_i(p)$, the term $Q_{\rm e^{\pm}}^{\rm 2nd}(p)$ refers to the source rate of secondary electrons e$^-$ and positrons e$^+$. For the best-fit scenario of the AGN corona, these secondaries emerge at the highest energies predominantly from so-called hadronic $\gamma$-rays (i.e.\ $\gamma$-rays that are produced by hadronic interactions.). In addition, a significant contribution from the second generation of secondaries---which are produced by the so-called secondary leptonic $\gamma$-rays (i.e.\ $\gamma$-rays that are produced by leptonic interactions of the secondaries.)---is present at low energies (see left panel of Fig.\ref{fig:had+lept}).

On the right panel of Fig.\ref{fig:had+lept} the solution of the transport equation (\ref{eq:transportEQ}) is shown for the best-fit scenario of the AGN corona of NGC 1068. In case of the non-thermal protons, the typical hard spectrum from SDA (see e.g.\ Ref.~\cite{walter2025stochastic} for more details) is obtained, with $n_{\rm p}^{\rm c}(E\lesssim 10^{13}\,\text{eV})\propto E^{1-\varkappa}$ at these energies where the acceleration is significantly faster than the energy losses ($\tau_{\rm acc}^{\rm c}\ll \tau_{\rm loss}^{\rm c}$), as well as the pile-up bump around the energy where those timescales are equal ($\tau_{\rm acc}^{\rm c}\sim \tau_{\rm loss}^{\rm c}$). The spectrum of non-thermal electrons is significantly softer, since primary electrons are not sufficiently accelerated by SDA ($\tau_{\rm acc}^{\rm c}\gg \tau_{\rm loss}^{\rm c}$), so that the overall spectrum consists only of the cooled secondaries.

The starburst regions of NGC\,1068 and NGC\,7469 show a ring-like structure on kiloparsec scales, which we also account in our model. Moreover, particles are most likely accelerated by supernova remnants (SNRs) via diffusive shock acceleration (DSA), so that the stochastic acceleration can be neglected. Consequently, we adopt that the momentum diffusion and energy gain term in Eq.~(\ref{eq:transportEQ}) vanish for the starburst regions and use the same two-zone approach as in the E+22 model.

The updated treatment of the non-thermal particle population is subsequently used to determine the resulting photon and neutrino flux from the individual emission sites such as in the E+22 model, except for the treatment of the $\gamma\gamma$-pair attenuation in the vicinity of the corona. 
In addition to the internal attenuation by the isotropic X-ray target distribution, the updated model can also account for: \emph{(i)} the impact of the anisotropic, external X-ray photons that radially leave the corona (\emph{int+ext att} scenario); and moreover, \emph{(ii)} a shrunken $\gamma$-ray production site with a spatial extension that is smaller than the adopted radius of the X-ray source (\emph{int+shield+ext att} scenario), so that the surrounding isotropic X-ray field further increases the $\gamma$-ray opaqueness.

\section{Methods}\label{sec:methods}

In the following, we use a Markov Chain Monte Carlo (MCMC) approach to study the parameter space of two potential high-energy neutrino sources and determine the best-fit solution. Hereby, current knowledge on the parameters is used as prior information. For the AGN-corona, the studied parameters are the scale radius $R^{\rm c}$ (related to the corona radius through $R \equiv {R^{\rm c}}R_{\rm s} $, where $R_{\rm s}$ is the Schwarzschild radius), the plasma beta $\beta^{\rm c} \equiv 8\pi n^{\rm c} k_{\rm B}T_{\rm gas}^{\rm c}/(B^{\rm c})^2$ (related to the magnetic field $B$), the optical thickness $\omega_{\rm T}^{\rm c}\equiv n^{\rm c}\sigma_{\rm T}R$ (related to the gas density $n^{\rm c}$) and fraction $f^{\rm c}$ of the gas density converting into CRs. The terms $k_{\rm B}$ and $\sigma_{\rm T}$ are the Boltzmann constant and the Thomson cross-section, respectively. For the starburst ring, the parameters are the magnetic field $B^{\rm s}$, the gas density $n^{\rm s}$, the fraction $f^{\rm s}$ of SN energy turned into CRs and the initial CR spectral index $q^{\rm s}$. The parameters priors are shown in purple in Fig.~\ref{fig:corner_1068}. 

For the MCMC fit the \emph{int+ext att} scenario is adopted with respect to the coronal $\gamma\gamma$-pair attenuation. In the post-processing also different scenarios are addressed.

\subsection{NGC\,1068}
\begin{figure}
    \centering
    \vspace{-.5cm}
    \includegraphics[width=0.98\linewidth]{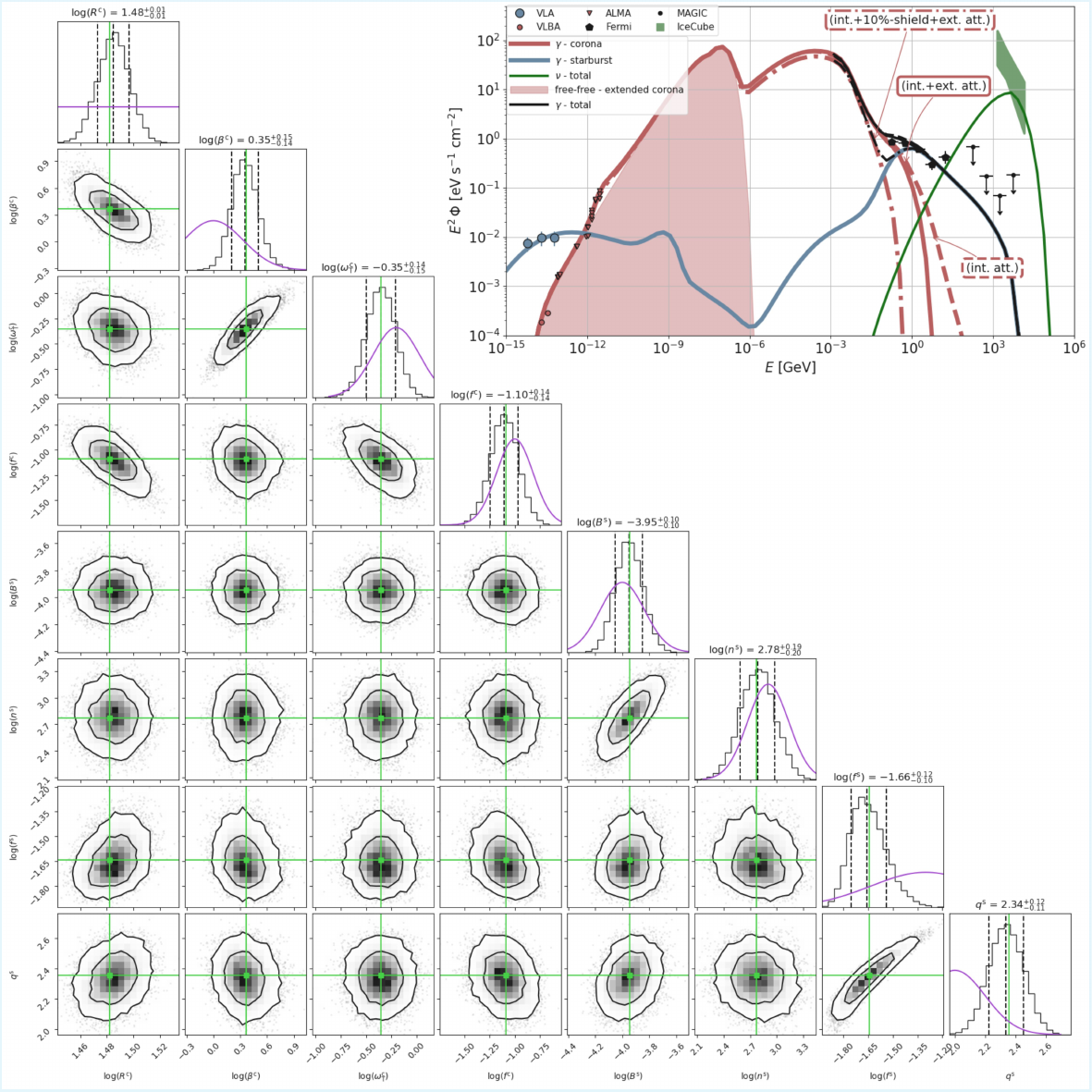}
    \caption{On the left, the parameters space explored through the MCMC for NGC$\,1068$: the two-dimensional and one-dimensional posteriors from the sampling are shown in black.
    In purple, the prior distributions are also shown. In green, the extracted values with the best likelihood are shown. On the right, the resulting model prediction (lines) and the used data (markers) of the SED of NGC 1068: The data consist of the VLA radio data \cite{wynn19853, gallimore2004parsec}, the sub-mm/mm data \cite{mutie2025consistent} as well as the \textit{Fermi}-LAT $\gamma$-ray data \cite{abdollahi2022incremental} and the MAGIC upper limits \cite{acciari2019constraints} at high energies. The green band marks the neutrino confidence band, provided by IceCube \citep{icecube2022evidence}.}
    \label{fig:corner_1068}
\end{figure}

The thermal radiation spectrum of NGC\,1068 is fixed such as in E+22 adopting an intrinsic coronal X-ray luminosity (between 2 and 10 keV) of $L_X = 7 \times 10^{43}\,\text{erg}/\text{s}$. 
The details about the assumed priors --- together with the corresponding references --- and the resulting best fit parameters for NGC\,1068 are shown in Fig.~\ref{fig:corner_1068}. The multi-wavelength data match the ones from E+22 except for some updated mm data points taken from Ref.~\cite{mutie2025consistent}. In the radio band we can partially resolve the different spatial emission zones. In addition to the corona and starburst, a diffuse emission extending up to $\sim$ pc scales --- the so called \textit{extended corona} region --- is included but fixed to explain the VLBA observations (see E+22 for details). We point out that our model accounts only for the non-thermal emission processes. At $\gamma$-ray energies the resolution is not sufficient to distinguish between corona and starburst, so that both environments can contribute to the signal.

\subsection{NGC\,7469}

The hints of high-energy neutrinos from the direction of NGC\,7469 with the absence of any $\gamma$-ray counterpart indicates that those neutrinos seem to originate from the AGN corona.
A 90\% confidence band for a possible true neutrino flux has been provided in Ref.~\citep{sommani2025two}. 
The necessary intrinsic X-ray target field (to absorb the $\gamma$-rays) is taken from Ref.~\cite{mehdipour2018multi} with an intrinsic coronal X-ray luminosity (between 2 and 10 keV) of $L_X = 1.7 \times 10^{43}\,\text{erg}/\text{s}$. Hereby, the \textit{Chandra}-HETGS, \textit{HST}-COS and \textit{Swift}-UVOT data from 2015 have been fitted yielding an X-ray flux that is just slightly larger than the results from previous epochs. 

The $\gamma$-ray upper limits were derived analyzing $15.5$\,years of \textit{Fermi}-LAT data in the energy range 0.1--1000\,GeV using \textsc{Fermipy}~\citep{fermipy}. We apply the same prior information on the corona in an MCMC approach to model the high-energy data. 

\section{Results}\label{sec:results}
\vspace{-.2cm}
\subsection{NGC\,1068}
Using the observational data of NGC\,1068 from radio to $\gamma$-ray frequencies that are associated to non-thermal emission signatures, the Fig.~\ref{fig:corner_1068} illustrates the resulting posterior distribution of the MCMC fit. Most of them are in close agreement with log-normal prior expectations and well constrained. 
In Fig.~\ref{fig:corner_1068} we show the results for the best fit parameters extracted from the MCMC procedure. The large-scale radio data are explained through secondary synchrotron emission from the starburst ring, whereas most of the small-scale radio data can be explained by free-free emission from the diffuse gas in the extended corona. But to explain the mm-bump (at $\sim 2 \times 10^{-12}$\,GeV) secondary synchrotron emission from the corona (with $R^{\rm c} = 30.2\pm 0.7\,R_{\rm s}$) is needed. To explain the \textit{Fermi}-LAT data, both the corona and the starburst ring are needed, providing $\gamma$-rays predominantly via secondary IC radiation and hadronic pion production, respectively. The high-energy neutrino emission via hadronic pion production that originates in the corona is in good agreement with the observations above about 10\,TeV. 
Moreover, it is shown that the external $\gamma\gamma$-pair attenuation has only a minor impact on the resulting sub-GeV $\gamma$-ray flux. But a shrunken $\gamma$-ray production site by 10\% already yields a considerable deficit that would enable an increased production of high-energy $\gamma$-rays and neutrinos.

\subsection{NGC\,7469}

\begin{figure}[htbp]
    \centering
    \includegraphics[width=0.48\linewidth]{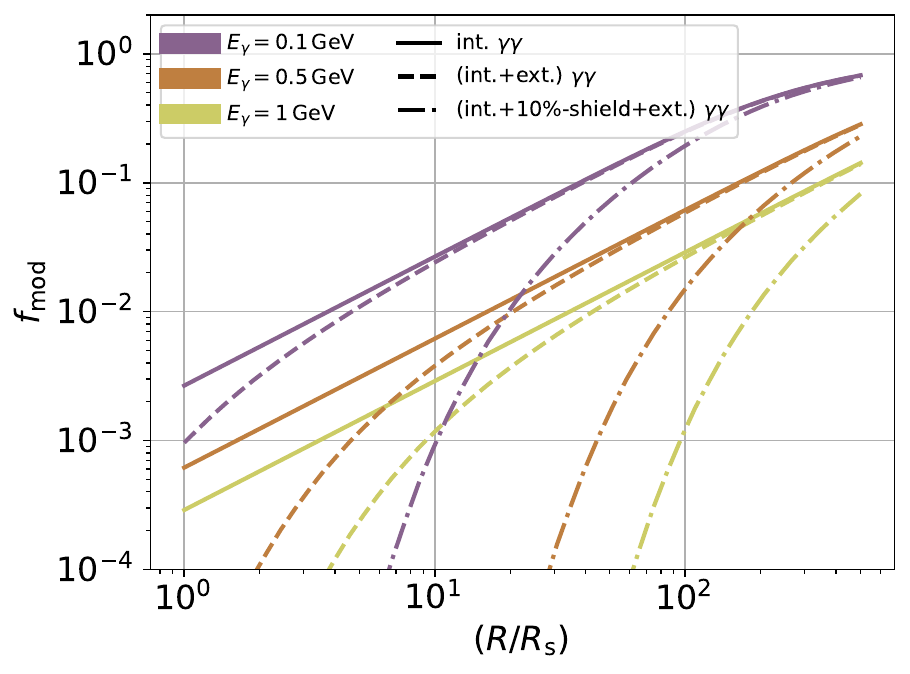}
    \includegraphics[width=0.51\linewidth]{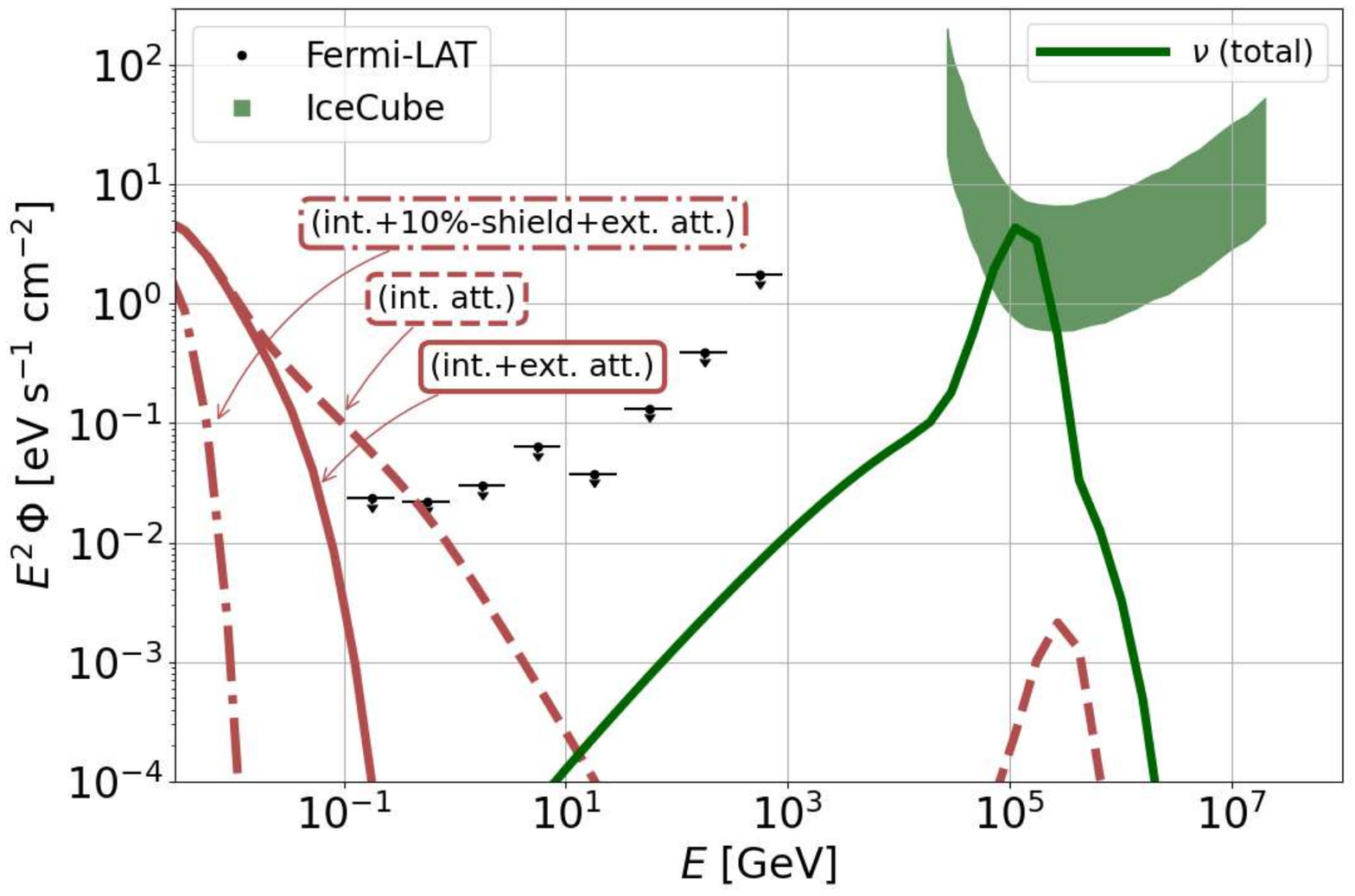}
    \caption{On the left, flux modification factor due to $\gamma\gamma$-pair attenuation for three different $\gamma$-ray energies (according to the colour) dependent on the size of the corona of NGC\,7469. On the right, high energy end of the resulting flux spectrum for NGC\,7469.}
    \label{fig:ngc7469} 
\end{figure}
Using the observational high-energy data of NGC\,7469 the MCMC fit predicts a bump of high-energy neutrinos from photomeson production at about about 100\,TeV, which is in good agreement with the current hints for neutrino emission, as shown in right Fig.~\ref{fig:ngc7469}. To not exceed the upper limits by \textit{Fermi}-LAT the \textit{int+ext att} scenario is needed as well as a small coronal radius of $R=5\,R_{\rm s}$. In more detail, the left Fig.~\ref{fig:ngc7469} exposes that the necessary attenuation of sub-GeV $\gamma$-rays by more than an order of magnitude can only be realized for $R^{\rm c}\lesssim 10 R_{\rm s}$. 

\section{Conclusions and Discussion} \label{sec:conclusions}

In this work, we update the spatially homogeneous, spherically symmetric, steady state two-zone model for the AGN-starburst composite galaxies firstly introduced in E+22. We aim to reproduce the multimessenger SED of two Seyfert galaxies, NGC\,1068 and NGC\,7469. Unlike E+22, the fit procedure is based on a Markov Chain Monte Carlo method, which allows us to use current knowledge on the AGN corona and starburst parameters as prior information. The three major changes in the model with respect to E+22 are associated with the AGN corona: \textit{(i)} A proper treatment of the stochastic acceleration process; \textit{(ii)} the leptonic cascades contribution originating from the first generation of hadronically produced $e^{\pm}$ secondaries; and \textit{(iii)} additional $\gamma\gamma$-pair attenuation scenarios. 

In the case of NGC\,1068, our best fit results indicate that both the starburst ring and the AGN corona are needed to explain the $\gamma$-ray emission---similar to the E+22 findings. But due to the proper treatment of the stochastic acceleration process, the high-energy neutrino spectra are significantly harder, showing a maximal energy flux at about $10\,\text{TeV}$. Hereby, a significantly lower coronal CR-to-thermal gas density pressure of $P_{\rm CR}/P_{\rm gas} = 0.07$ is needed. The posterior distributions of the parameters are quite similar to the prior distributions and most parameters can be well constrained. Moreover, it is shown that the details on the spatial distribution of the X-ray photon target and the $\gamma$-ray production site, respectively, has a significant impact on the results. 
Thus, a scenario where the $\gamma$-ray production site is also shielded by the X-ray photon target is expected to enable an improved fit to the neutrino data below 10\,TeV.
In the case of NGC\,7469, an explanation of the high-energy emission is only possible for a small corona radius ($\lesssim 10 R_{\rm s}$), even considering the external $\gamma\gamma$-pair attenuation. 
Future examinations need to clarify whether these coronal conditions also enable an explanation of the mm-bump (similar to NGC$\,1068$) indicated by ALMA observations (in prep.). 
Since NGC\,7469 is a type 1.5 Seyfert galaxy that shows X-ray variability, there is some uncertainty on the adopted intrinsic X-ray target field. However, we have already used a model of this target in its high state, so that a significantly thicker X-ray density, and hence, an increased $\gamma$-ray opaqueness, is not expected. 

\section*{Acknowledgements}
BE, SS, and JBT acknowledge support from the Deutsche Forschungsgemeinschaft (DFG): this work was performed in the context of the DFG-funded Collaborative Research Center SFB1491 "Cosmic Interacting Matters - From Source to Signal", project no. 445052434.
S.d.P. acknowledges support from ERC Advanced Grant 789410.

\vspace{-3mm}
\begingroup
\footnotesize
\bibliographystyle{JHEP}
\setlength{\bibsep}{1pt}
\bibliography{references}

\end{document}